\documentclass[RNAAS,twocolumn]{aastex62}
\usepackage{newtxtext}
\usepackage[slantedGreek]{newtxmath}
\makeatletter
\long\def\frontmatter@title@above{
\vspace*{-\headsep}\vspace*{\headheight}
\footnotesize\noindent{\sc \today}\\[2pt]
{\footnotesize Typeset using \LaTeX\ {\bf RNAAS} style in AASTeX62}
\par\vspace*{-\baselineskip}\vspace*{0.625in}}
\makeatother
\begin{document}

\title{Arcminute MicroKelvin Imager observations at 15~GHz of the
2020 February outburst of Cygnus X-3}

\correspondingauthor{David A.\ Green}
\email{dag@mrao.cam.ac.uk}

\author[0000-0003-3189-9998]{David A.\ Green}
\affiliation{Cavendish Laboratory,\\
             19 J.~J.\ Thomson Ave., Cambridge, CB3 0HE, UK}

\author{Patrick Elwood}
\affiliation{Cavendish Laboratory,\\
             19 J.~J.\ Thomson Ave., Cambridge, CB3 0HE, UK}

\vskip1cm

\keywords{High mass x-ray binary stars (733) --- Galactic radio sources
(571) --- Radio continuum emission (1340) --- Variable radiation sources
(1759)}

\vskip1cm

\section{}

The high mass X-ray binary \object{Cygnus X-3} shows occasional giant
radio flares, with flux densities up to $\sim 10$~Jy. The first such
burst was observed in 1972 \citep{1972Natur.239..440G}, and subsequently
there have been several similar flares, e.g.\ see
\citet{1986ApJ...309..707J, 1995AJ....110..290W, 1997MNRAS.288..849F,
2001ApJ...553..766M, 2012MNRAS.421.2947C, 2016MNRAS.456..775Z,
2017MNRAS.471.2703E} for some examples.

Here we report observations of a recent giant radio flare from Cygnus
X-3 in 2020 February, made with the Arcminute MicroKelvin Imager (AMI,
\citealt{2008MNRAS.391.1545Z, 2018MNRAS.475.5677H}). These observations
were triggered by the detection of an increase in the emission from
Cygnus X-3 seen at 37~GHz with the Mets\"ahovi Radio Observatory (Karri
Koljonen, private communication). The observations were made with the
AMI `Small Array' which is a radio interferometer consisting of ten
3.7-m diameter antennas. A single linear polarisation, Stokes parameter
$I+Q$, was observed over a frequency range of 13.0 to 18.0~GHz. However,
in practice the ends of this frequency range were not used (either due
to poor sensitivity or, at the lower frequencies, satellite
interference), and the observed band was 14.2 to 17.4~GHz.

\begin{figure*}[htb!]
\centerline{\includegraphics[width=17.5cm]{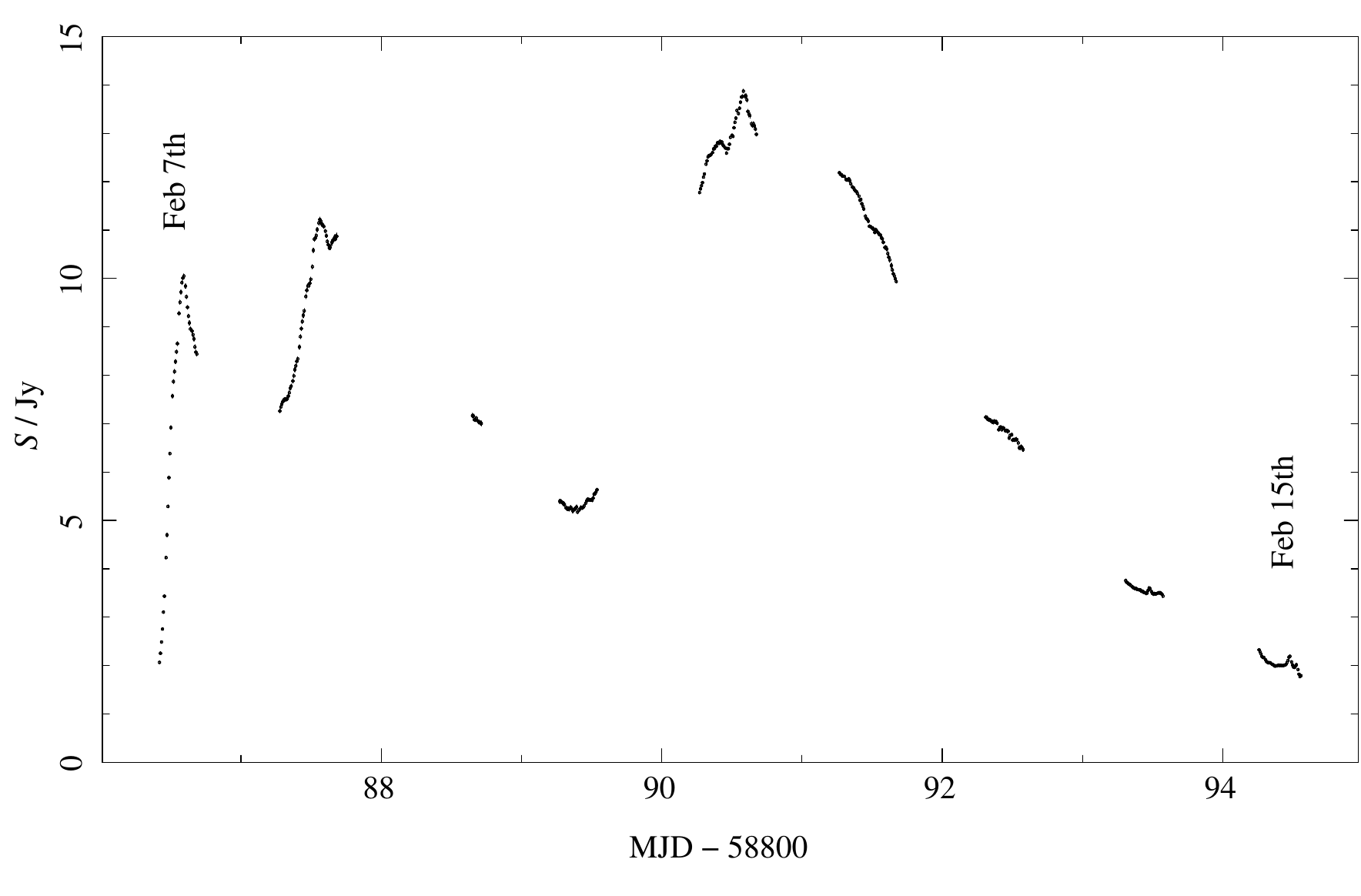}}
\caption{AMI Small Array observations of Cygnus X-3 during the period
2020 February 7th to 15th, at 15 GHz (Stokes $I+Q$). Each data point is
a 10-min average. Statistical error bars are plotted, although these are
usually smaller than the size of symbols.\label{fig:ami-sa}}
\end{figure*}

The observations consisted of 1-hour integrations on Cygnus X-3,
interleaved with 400-s observations of a nearby, compact calibrator
source J2052$+$3635. The first observation was made on 2020 Feb 7th, for
$\approx 6.4$ hours. Long observations (up to $\approx 9.7$ hours, after
flagging) were made on all but one of the following days up to and
including 2020 Feb 15th; the exception was Feb 9th, when only a short
observation ($\approx 1.5$ hours) was possible, due to extremely high
winds, which meant the array had to be stowed. The number of antennas
available during these observations was usually 8 (except on Feb 9th and
13th when the number was, respectively, 7 and 9).

The data were processed using standard procedures, with the flux density
scale established from short observations of the standard calibrator
source 3C286 which were made most days, together with the `rain gauge'
measurements made during the observations which were use to correct for
varying atmospheric conditions (see \citeauthor{2008MNRAS.391.1545Z}).
The data were flagged: (i) automatically to eliminate bad data due to
various technical problems, interference, and when some antennas were
shadowed at the end of the observations (at low elevations); (ii)
manually, to eliminate remaining interference, and some periods with
very heavy rain. The interleaved observations of J2052$+$3635 provided
the initial phase calibration of each antenna in the array throughout
each observation. The amplitudes of the J2052$+$3635 observations were
used to check the consistency of the flux density scale during the
observations. It was found that the r.m.s.\ deviation of the
J2052$+$3635 flux densities was $1.8$~per~cent. Subsequently the
observations were phase self-calibrated on a timescale of 10 min. Flux
densities were derived for 10-min averages, for 5 broad frequency
channels covering 14.2 to 17.4~GHz, and then a power law fit was made to
obtain a flux density at 15~GHz.

Figure~\ref{fig:ami-sa} shows the 15-GHz light curve of Cygnus X-3 from
these observations. This shows a rapid increase on Feb 7th, rising from
$\approx 2$~Jy to a peak of $\approx 10$~Jy in just over 4 hours. During
this period the radio spectrum across the AMI band changed from
optically thick (flux density rising with increasing frequency) to
optically thin (flux density falling with increasing frequency)
synchrotron emission. In subsequent days Cygnus X-3 showed other peaks
at $\approx 11$~Jy on Feb 8th, and $\approx 14$~Jy on Feb 11th, after
which it faded, approximately exponentially (with a characteristic time
of ~1.7 days).

\acknowledgments
We thank the staff of the Mullard Radio Astronomy Observatory,
University of Cambridge, for their support in the maintenance, and
operation of AMI. We acknowledge support from the European Research
Council under grant ERC-2012-StG-307215 LODESTONE.

%------------------------------------------------------------------------------%
\end{document}